\documentclass[10pt,conference]{IEEEtran}
\input{psfig.sty}
\input{epsf.sty}

\usepackage{hyperref} 
\usepackage{fancybox}   
\usepackage{psfig}      
\usepackage{graphicx}
\usepackage{amsmath}
\usepackage{color}
\usepackage{epsfig}
\usepackage{amssymb}
\usepackage{verbatim}
\usepackage{enumitem}
\usepackage{bbm}

\usepackage[T1]{fontenc}
\usepackage[latin9]{inputenc}
\usepackage{color}
\usepackage{tikz}
\usepackage{float}
\usepackage{amssymb,amsmath,amsfonts}
\usepackage{mathrsfs}
\usepackage{amsgen,amsfonts,amsbsy,amsthm}
\usepackage{algorithmic,algorithm}
\usepackage{amsmath}
\usepackage{graphicx}
\usepackage{setspace}
\usepackage{amssymb}
\usepackage{esint}
\usepackage{bbm}
\usepackage{amsmath}
\usepackage{graphicx}
\usepackage{setspace}
\usepackage{amssymb}
\usepackage{esint}
\usepackage{subcaption}
\usepackage{url}

\IEEEoverridecommandlockouts

\newcommand{\remove}[1]{}

\usepackage{setspace}

\begin{document}

\title{Impromptu Deployment of Wireless Relay
  Networks:  Experiences Along a Forest Trail 
\thanks{Arpan Chattopadhyay, Avishek Ghosh, Bharat Dwivedi,
S.V.R. Anand, and Anurag Kumar are with the Department of ECE, Indian Institute of Science, Bangalore, India; email: arpanc.ju@gmail.com,
\{avishek, bharat, anand, anurag\}@ece.iisc.ernet.in.}
\thanks{Akhila S. Rao is with KTH, Royal Institute of Technology, Stockholm, Sweden; email: akhila@kth.se.}
\thanks{Marceau Coupechoux is with Telecom ParisTech and CNRS LTCI, 
Dept. Informatique et Reseaux, 23, avenue d'Italie, 
75013 Paris, France; email: marceau.coupechoux@telecom-paristech.fr.}
 \thanks{This work was supported by (i) the Department of 
Electronics and Information Technology (India) and the National Science Foundation (USA) via an  
Indo-US collaborative project titled Wireless Sensor Networks for Protecting Wildlife and Humans, and 
(ii) the J.C. Bose National Fellowship (of the Govt. of India).}
}


\newcounter{one}
\setcounter{one}{1}
\newcounter{two}
\setcounter{two}{2}

%

\vspace{-3mm}

\author{
Arpan~Chattopadhyay, Avishek Ghosh, Akhila S. Rao, 
Bharat Dwivedi,\\ S.V.R. Anand, Marceau~Coupechoux, 
and Anurag~Kumar\\
\vspace{-1.5cm}
}

\maketitle
\thispagestyle{empty}

\vspace{-18mm}

\begin{abstract}
  We are motivated by the problem of impromptu or as-you-go deployment
  of wireless sensor networks.  As an application example, a  person, starting from a sink node, 
  walks along a forest trail, makes link
  quality measurements (with the previously placed nodes) at equally spaced
  locations, and deploys relays at some of these locations, so as to
  connect a sensor placed at some a priori unknown point on the trail
  with the sink node.  In this paper, we report our experimental experiences with
  some as-you-go deployment algorithms.  Two algorithms are based on
  Markov decision process (MDP) formulations; these
  require a radio propagation model.  We also study purely measurement based
  strategies: one heuristic that is motivated by our MDP formulations, 
one asymptotically optimal learning algorithm, and
  one inspired by a popular heuristic. We extract a statistical
  model of the propagation along a forest trail from raw measurement data, 
  implement the algorithms experimentally in the forest, and compare
  them.  The results
  provide useful insights regarding the choice of the deployment
  algorithm and its parameters, and also demonstrate the necessity of
  a proper theoretical formulation.
\end{abstract}

\vspace{-4.5mm}
\section{Introduction}\label{section:introduction}
\vspace{-4mm}
Our work in this paper is motivated by 
the need for as-you-go deployment of wireless relay networks over large terrains, 
such as forest trails, where planned deployment would be  
time-consuming and difficult. As-you-go deployment is the only choice when the network is temporary and needs
to be quickly redeployed, or when the
deployment needs to be stealthy. 

\begin{figure}[!t]
\begin{center}
\includegraphics[height=1.5cm, width=8cm]{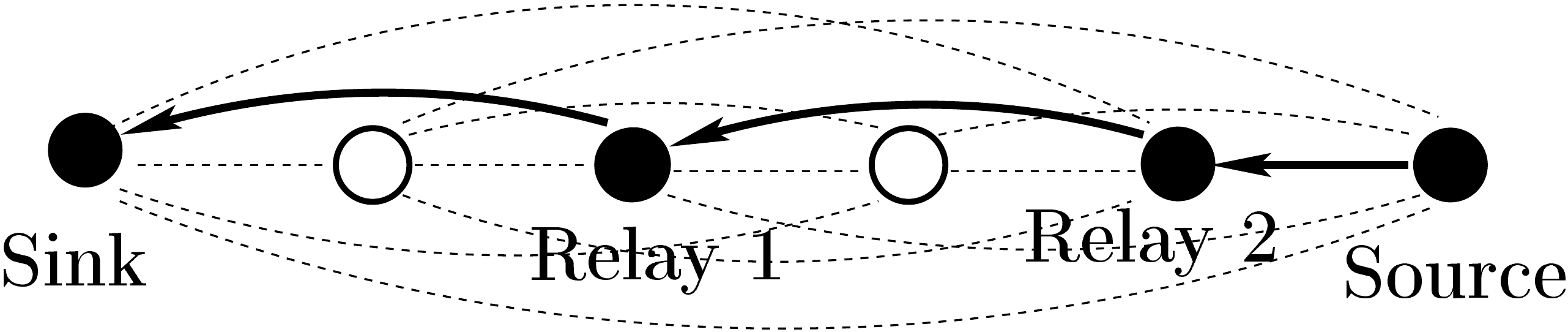}
\end{center}
\vspace{-4mm}
\caption{Two wireless relays  (filled dots) deployed along a line to connect a source 
to a sink by a multihop path. The unfilled dots show other potential relay placement locations, 
the thin dashed lines indicate all the potential links between the potential placement 
locations, and the solid lines with arrowheads indicate the links actually used in the deployed network.}
\label{fig:line-network-general}
\vspace{-6mm}
\end{figure}

In this paper, we are concerned with an {\em experimental study} of the problem of deploying wireless
relay nodes as a deployment agent walks along a forest trail, in order to 
connect a sink at the start of the trail to a sensor
that would need to be deployed at an a priori unknown point. The sensor could be an animal activity detector
based on passive infra-red (PIR) sensors, or even a ``camera trap,'' that has to
be placed near a watering hole that is known to exist somewhere just
off the trail. Figure~\ref{fig:line-network-general} depicts the abstraction of the problem along a line. 
The sink has a  ``backhaul'' communication link  to a control center.

\vspace{-1mm} 

In  planned deployment, we need to place relay nodes at \emph{all} potential locations (see Figure~\ref{fig:line-network-general} 
 for a simple depiction) 
and  measure the 
qualities of all possible (solid and dotted) links (between all pairs of potential locations; see 
Figure~\ref{fig:line-network-general}) 
in order to decide where to place the relays. 
This would yield a global optimal solution, but with huge time and effort. 
With as-you-go 
deployment, the next relay
placement locations depend on the radio link qualities to the
previously placed nodes; link qualities and the location of the
sensor node are discovered as the agent walks along the trail. 
Such an approach 
requires fewer measurements compared to planned deployment, but is suboptimal.  
In this paper, we report the results of our experimentation with
some as-you-go deployment algorithms (taken from our prior work
\cite{chattopadhyay-etal13backtracking-as-you-go_arxiv} which is an extension of 
\cite{chattopadhyay-etal13measurement-based-impromptu-placement_wiopt}, one heuristic adapted
from the literature, and one proposed in this paper) on a forest trail in the campus of  Indian Institute of
Science, Bangalore.

\emph{Related Work:} Souryal et al. \cite{souryal-etal07real-time-deployment-range-extension} provide a
deployment protocol employing real-time assessment of wireless link quality. As the agent
walks away from the sink, signal quality measurements are made with
already placed nodes, and a threshold-based algorithm determines when
it is opportune to place another relay. One of the algorithms that we
include in our experimental study is a simple adaptation of the
algorithm reported in \cite{souryal-etal07real-time-deployment-range-extension}. Liu et al. \cite{liu-etal10breadcrumb}
proposed the Breadcrumb System to aid fire-fighters inside a
building. The system drops tiny radio relays (``breadcrumbs'') as a
firefighter walks; the impromptu wireless network so created is used 
to send the firefighter's physiological parameters back to the 
control truck outside the
building. 
See \cite{aurisch-tlle09relay-placement-emergency-response} and
\cite{howard-etal02incremental-self-deployment-algorithm-mobile-sensor-network}
for other approaches.

However, there has been little effort to rigorously formulate the
problem in order to derive optimal policies which are insightful and
which can provide improved performance compared to reasonable
heuristics.  Recently, Sinha et
al. \cite{sinha-etal12optimal-sequential-relay-placement-random-lattice-path}
have provided a Markov decision process (MDP) formulation for
establishing a multi-hop network between a destination and an unknown
source location by placing relay nodes along a random lattice
path. They assume a given deterministic mapping between power and
wireless link length; this is achieved by employing a very
conservative fade margin to take care of the joint effects of
shadowing and fast fading, thereby to determine the transmit power
required to maintain the link quality over links of a given length. We
considered the possibility of link outage, and brought in the idea of
\emph{measurement based} optimal impromptu placement, first in
\cite{chattopadhyay-etal13measurement-based-impromptu-placement_wiopt}
and next in the extended version
\cite{chattopadhyay-etal13backtracking-as-you-go_arxiv}.

\vspace{-5mm}
\section{System Model and Deployment Setting}\label{section:system_model_and_deployment_setting}
\vspace{-2mm} Deployment is done by a single agent (in one ``pass'') along a line
discretized in steps of length $\delta$ (e.g., $20$~meters), and these
discrete locations are the potential relay locations. 

\vspace{-4mm}
\subsection{Channel Model}\label{subsection:channel_model}
\vspace{-2mm} 

The received power (for the $k$-th packet) for a link
 of length $r$ is given by $ P_{rcv,k}=P_T c (\frac{r}{r_0})^{-\eta}H_k W $,  
where $P_T$ is the transmit power, $c$ corresponds to the path-loss at
the reference distance $r_0$, $\eta$ is the path-loss exponent, $H_k$
denotes the fading random variable (varying with time for a link) for
the $k$-th packet, and $W$ denotes the shadowing random variable
(constant for a link but varies over different links).  $W$ is  modeled as a log-normal random variable;
$W=10^{\frac{\nu}{10}}$, $\nu \sim \mathcal{N}(0, \sigma^2)$.  The
mean received power (averaged over fading) in a link with shadowing
realization $w$ is $\overline{P}_{rcv}=P_T c (\frac{r}{r_0})^{-\eta}w
\mathbb{E}(H)$. {\em We assume that the shadowing random
  variables at any two different links are independent; this holds if
  $\delta$ is greater than the shadowing decorrelation distance (see Section~\ref{section:virtual_walking_experimental_results}).}

A link is considered to be in \emph{outage} if the received signal
power (RSSI) drops (due to fading) below a value $P_{rcv-min}$. For packet size of $140$~bytes  and for TelosB motes, 
the packet error rate  
(PER) is 
less than $2\%$ at RSSI $-88$~dBm, and increases rapidly as RSSI decreases
further (see \cite{bhattacharya-etal13smartconnect-comsnets}).  
{\em Hence, for TelosB motes, we choose $P_{rcv-min}=-88$~dBm.}

The transmit power of each node can be chosen from a discrete set,
$\mathcal{S}$.  If the fading statistics
are known, then the outage probability is
$P_{out}(r,\gamma,w):=\mathbb{P}(\gamma c (\frac{r}{r_0})^{-\eta}H w
\leq P_{rcv-min})$ for a link of length $r$ and shadowing realization
$w$ ($w$ is unknown) at a particular transmit power level $\gamma$.
We can measure $P_{out}(r,\gamma,w)$ for a link with length $r$,
transmit power $\gamma$ and shadowing realization $w$ by sending a
large number of packets over multiple coherence times and then obtaining the fraction of packets
whose RSSI value is below
$P_{rcv-min}$.

\vspace{-2mm}
\subsection{Deployment Process }\label{subsection:deployment_process_notation}
\vspace{-4mm}

\begin{figure}[!t]
\begin{center}
\includegraphics[height=1.4cm, width=9cm]{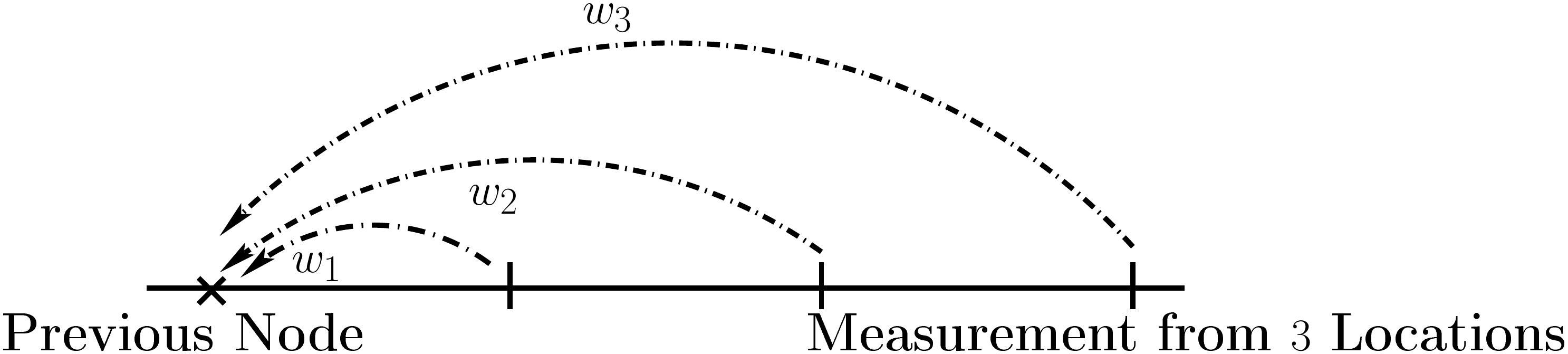}
\end{center}
\vspace{-4mm}
\caption{The agent places a relay (shown by an ``$\mathsf{x}$'') and makes 
measurements to obtain the outage probabilities $\{p_{out}(r,\gamma,w_r)\}$, $r \in \{1,2,3\}$, $\gamma \in \mathcal{S}$,  
from successive potential locations. With $B=3$, 
 measurements from $3$ successive locations are made before making a placement decision.}
\label{fig:exploring_illustration}
\vspace{-7mm}
\end{figure}

We consider two deployment approaches: (i) {\em limited exploration based}, and 
(ii) {\em pure as-you-go}. 
We explain these alternatives with reference to Figure~\ref{fig:exploring_illustration}. The agent walks away from the sink 
(from left to right in the figure), evaluating whether to place relays at the \emph{potential} 
placement points that are at multiples  of the step length $\delta$. Suppose a relay has been placed at 
the position marked by the $\mathsf{x}$ in Figure~\ref{fig:exploring_illustration}.  
Let us denote by $w_r$, the realization of shadowing from the location which is 
$r$ steps ahead of the placed relay, to this relay. For deployment with {\em limited exploration}, 
as the deployment agent walks along the line, after placing a node, 
he measures the outage probabilities $P_{out}(r, \gamma, w_r)$ to the previous node from locations 
$r \in \{1, \cdots,B\}$, at each power level $\gamma \in \mathcal{S}$ (see Figure~\ref{fig:exploring_illustration}; 
$\mathcal{S}$ is the finite set of transmit power levels that a node can use). 
Then he places the relay at one of the locations $1,2, \cdots,B$, sets it to operate at a certain transmit 
power (both decisions being made by the algorithm). Recursively, this relay then becomes $\mathsf{x}$, 
the deployment agent moves forward and applies the same procedure to deploy additional relays. 
If the source location is encountered within $B$ steps from the previous node 
(i.e., the "current $\mathsf{x}$"), 
then the source is placed.

With {\em pure as-you-go (no exploration)}, the agent 
measures $\{P_{out}(r, \gamma, w_r)\}_{ \gamma \in \mathcal{S}}$ 
when he is $r$ steps away from the previous relay, and at each step (after making the measurement) 
the algorithm decides whether to 
place a relay there or whether to advise the agent to move on without placing. 
In this process, if he has walked $B$ steps away from the previous 
relay, or if he encounters the source location, then he must place a node.

\vspace{-3mm}
\subsection{Traffic Model}\label{subsection:traffic_model}
\vspace{-3mm} 

The formulations from which the algorithms are derived in
\cite{chattopadhyay-etal13backtracking-as-you-go_arxiv} assume a very
light traffic model, the assumption being that there is only one
packet in the network at a time; we call this the ``lone packet
model.''  Thus, the formulations assume that there are no simultaneous
transmissions to cause interference. It is a good approximation for 
sensor networks that 
just carry an occasional alarm packet, or low duty-cycle measurement packets. It has been shown in \cite{bhattacharya-kumar12qos-aware-survivable-network-design} that, under a CSMA MAC, in order to achieve a target delivery 
probability under any packet arrival rate, it is necessary to 
achieve the target delivery probability under the lone packet traffic model.

\vspace{-3mm}
\subsection{Network Cost and Optimality Objective}\label{subsection:network_cost}
\vspace{-3mm}
Under lone packet traffic, the cost of a deployed network is the sum of certain hop costs. 
In case all the nodes have {\em wake-on radios,} the nodes normally stay in a very low current sleep mode.   
When a node has a packet, it sends a wake-up tone to the intended receiver. 
The receiver wakes up and the sender transmits the 
packet. The receiver sends an ACK packet in reply. 
Clearly, the energy spent in transmission and reception of data packets 
govern the lifetime of a node, given that the ACK size is negligible compared to packet size. 

We call the sink as node $0$, and the source as node $(N+1)$ ($N$ is the number of relays).  
We denote the transmit power of node~$i$
by $\Gamma_i$, and the outage probability in the link $(i,i-1)$ by
$P_{out}^{(i,i-1)}$.  The power required in the node electronics to
carry out the functions of reception is denoted by $P_r$.  

We use the
sum outage probability $\sum_{i=1}^{N+1}P_{out}^{(i,i-1)}$ as our
measure of end-to-end path quality. One motivation for this measure is
that, for small values of $P_{out}$, the sum-outage is approximately
the probability that a packet from the source encounters an outage
along the path. It can be argued that the rate of replacement of
batteries in the network is proportional to $\sum_k \Gamma_k$. 
Let $\xi_o$ denote the cost multiplier for outage and $\xi_r$ denote the cost
of a relay. Hence, a suitable cost of the network is $\sum_{i=1}^{N+1} \Gamma_i + \xi_o \sum_{i=1}^{N+1}P_{out}^{(i,i-1)}+ \xi_r N $.  
A {\em deployment policy} $\mu$, at each placement decision point, looks at all the past measurements and decisions, 
and provides the deployment agent with a placement decision. 
Let us denote by $N_x$ the number of relays deployed up to $x$~steps under a deployment policy 
$\mu$. Note that, $N_x $  
is a random variable where the randomness comes from the shadowing in the links encountered in the 
deployment process up to distance $x$. 
The objective is to find the Markov stationary optimal policy which minimizes the average cost per step:
\vspace{-3mm}

\footnotesize
\begin{equation}
 \mu^*:= \arg \min_{\mu} \limsup_{x \rightarrow \infty} \frac{\mathbb{E}_{\mu} 
( \sum_{i=1}^{N_x} \Gamma_i + \xi_o \sum_{i=1}^{N_x}P_{out}^{(i,i-1)}+ \xi_r N_x )}{x}\label{eqn:average_cost_objective}
\end{equation}
\normalsize

We can motivate the cost objective in (\ref{eqn:average_cost_objective}) as the relaxed version of the problem where we seek to 
minimize the mean transmit power per step, subject to a constraint on the mean outage per step and a 
constraint on the mean number of relays per step. $\xi_o$ and $\xi_r$ are Lagrange multipliers whose 
unit is mW. It follows from standard MDP theory that if the constraints are met with equality under a policy which is 
the optimal policy for (\ref{eqn:average_cost_objective}) for a given $(\xi_o,\xi_r)$, then that policy 
is optimal for the constrained problem also.

 \vspace{-5mm}
\section{Deployment Algorithms}\label{section:as_you_go_deployment_algorithms}
\vspace{-3mm}

\vspace{-1mm}
\subsection{An Optimal Algorithm with Limited Exploration (OptExploreLim)}\label{subsection:opt-back-description}
\vspace{-2mm}
Recall the deployment process in 
Section~\ref{subsection:deployment_process_notation} and the deployment objective 
in Section~\ref{subsection:network_cost}. 
Let us denote the optimal average cost per step by $\lambda^{*}$. 
Starting from the sink, or from a just placed relay, the  optimal policy $\mu^*$, given 
the measurements $P_{out}(u,\gamma,w_u)$, for all $u \in \{1,2,\cdots,B\}$ and $\gamma \in \mathcal{S}$, 
outputs the distance $u^*$ (in steps of size $\delta$) 
(at which the next relay 
has to be placed) and its transmit power $\gamma^*$. It was shown in 
\cite{chattopadhyay-etal13backtracking-as-you-go_arxiv} using an average 
cost Semi-Markov Decision Process (SMDP) formulation 
that:

\vspace{-2mm}
\footnotesize
\begin{equation}
 (u^*,\gamma^*)= \arg  \min_{u \in \{1,\cdots,B\}, \gamma \in \mathcal{S}}
(\gamma + \xi_o P_{out}(u,\gamma,w_u) + \xi_r -\lambda^{*}u) \label{eqn:smdp-optimal-policy}
\end{equation}
\normalsize

\vspace{-6mm}
\subsection{A Heuristic Algorithm with Limited Exploration (HeuExploreLim)}\label{subsection:heu-back-description}
\vspace{-4mm}
Starting from the sink or a just placed relay, and given the measurements $P_{out}(u,\gamma,w_u)$, for all $u \in \{1,2,\cdots,B\}$ and $\gamma \in \mathcal{S}$, 
this algorithm obtains the deployment distance $u$ and the node power $\gamma$ as follows:
\begin{equation}
\min_{u \in \{1,\cdots,B\}, \gamma \in \mathcal{S}} \frac{\gamma+\xi_o P_{out}(u,\gamma,w_u)+\xi_r}{u}
\label{eqn:heu-back-policy}
\end{equation}

{\em Remark:} This purely on-line heuristic optimizes an objective different
from OptExploreLim (see \cite{chattopadhyay-etal13backtracking-as-you-go_arxiv}), and is suboptimal 
for the Problem~(\ref{eqn:average_cost_objective}).

\vspace{-4mm}
\subsection{An Optimal Learning Algorithm with Limited Exploration (OptExploreLimLearning)}
\label{subsection:opt-explore-lim-learning-description}
\vspace{-4mm}

This stochastic approximation based algorithm  provides the same average cost as OptExploreLim.  The deployment
agent starts with an initial value $\lambda_0$, and places the first
relay (using the outage probabilities from $B$
locations for different transmit power levels) using the algorithm in
Equation~(\ref{eqn:smdp-optimal-policy}) (with $\lambda^*$ of (\ref{eqn:smdp-optimal-policy}) being replaced 
by $\lambda_0$). After placing the $(k+1)$-st
relay (using $\lambda_k$ in (\ref{eqn:smdp-optimal-policy})), we set $\lambda_{k+1}$ to be the actual average cost per step 
from the $(k+1)$-st relay to the sink node. It
can be shown that $\lambda_k \rightarrow \lambda^*$ with probability
$1$.

\vspace{-4mm}
\subsection{Optimal As-You-Go (OptAsYouGo) Algorithm}\label{subsection:opt-noback-description}
\vspace{-3mm}
It was shown in \cite{chattopadhyay-etal13backtracking-as-you-go_arxiv} that in the pure as-you-go case, 
it is optimal to place a relay at a distance $1 \leq r \leq (B-1)$ from the previous node if and only if  
$\min_{\gamma \in \mathcal{S}} (\gamma+\xi_o P_{out}(r,\gamma,w_r)) \leq c_{th}(r)$, and choose 
the minimizer as the transmit power. The 
threshold $c_{th}(r)$ is calculated by a value iteration arising out of an MDP formulation.  
$c_{th}(r)$ increases with $r$; since the outage probability, for a given $\gamma$ and $w$, 
increases with $r$, the chance of getting a link 
with small cost decreases as $r$ increases.

\vspace{-4mm}
\subsection{A Simple As-You-Go Heuristic (HeuAsYouGo)}\label{subsection:sim-heu-description}
\vspace{-3mm}
This is a modified version of the deployment algorithm 
proposed in \cite{souryal-etal07real-time-deployment-range-extension}. 
The power used 
by the relays is set to some fixed value, and at each potential location, the deployment agent  
checks whether the outage to the previous relay meets a pre-fixed target. 
After placing a relay, the next relay is placed at the last location  where the target 
outage is met; or place at the first location (after the previously placed relay) in the unlikely situation where the target outage is violated in the very 
first location itself. If the agent reaches the $B$-th step, 
he must place the next relay. This 
requires the deployment agent to move back by one step and place in case the outage target is violated for the first 
time in the second step or beyond. 
In practice, the transmit power might be chosen according to a lifetime constraint of the 
nodes, and the outage target can be chosen according to the mean outage per step constraint and the mean number of relays 
per step constraint. In order to make a fair comparison with OptExploreLim, we use the mean power 
(resp., mean outage) per link of 
OptExploreLim as the node transmit power (resp., the target outage) in HeuAsYouGo.

{\em Remark:} For any pair $(\xi_o, \xi_r)$, the OptExploreLim and OptAsYouGo algorithms 
require a statistical model of the channel in order to calculate $\lambda^*$ and $c_{th}(r)$. But 
HeuExploreLim, OptExploreLimLearning and HeuAsYouGo do not require any channel model; 
they are, therefore, the most useful in practice.

  \begin{figure}[t!]
\begin{centering}
\begin{center}
\includegraphics[height=3cm, width=4.2cm]{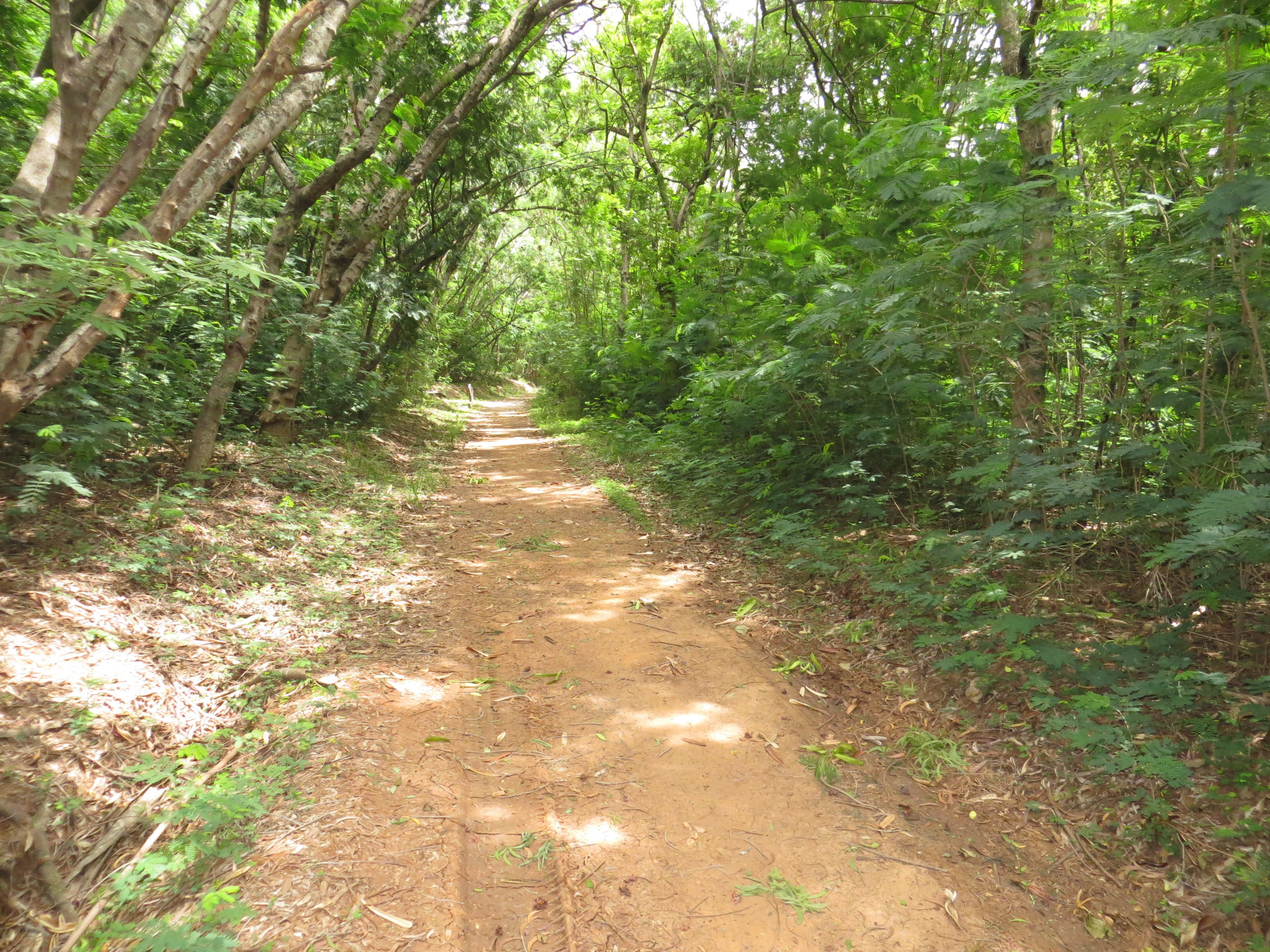}
\includegraphics[height=3cm, width=4.2cm]{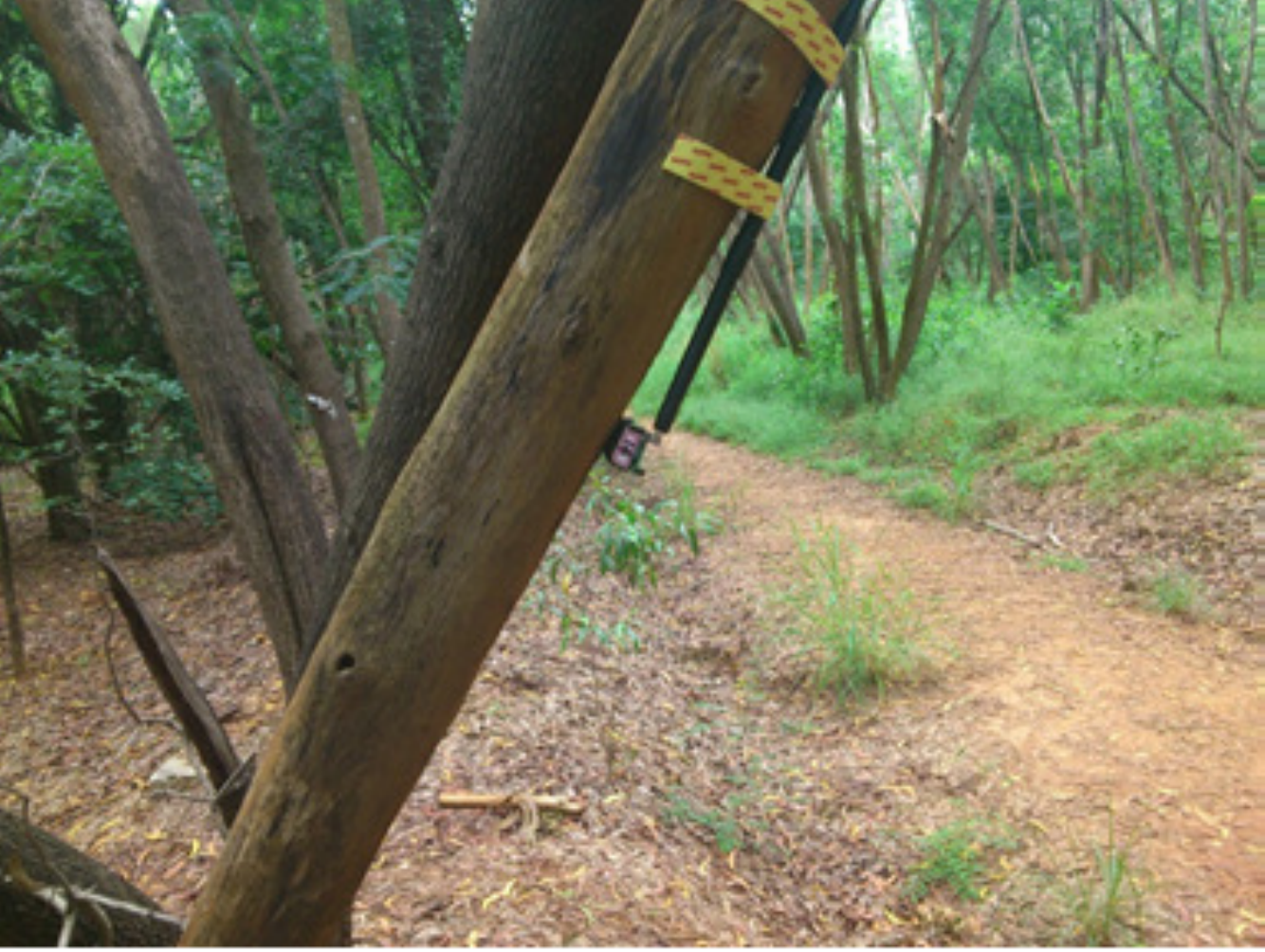}
\end{center}
\end{centering}
\vspace{-3mm}
\caption{Left: a segment of the trail along which deployment experiments were performed. Right: 
the wireless nodes (TelosB motes) were latched to the trees as shown, so that all the nodes were on 
one side of the trail, and the line of sight between the relays passed through the foliage.}
\label{fig:trail_image}
\vspace{-5mm}
\end{figure}

\vspace{-3mm}
\section{Radio Propagation Modeling}\label{section:radio_propagation_modeling}
\vspace{-3mm}
All our experiments were conducted on a trail in the forest-like Jubilee Gardens 
in the Indian Institute of Science campus (see Figure~\ref{fig:trail_image}). 
Our experiments were conducted  by placing the wireless devices on the edge of the trail 
so that the line-of-sight between the nodes passed through the foliage.
\vspace{-4mm}
\subsection{Modeling of Path-Loss and Shadowing}\label{subsection:shadowing_modeling}
\vspace{-2mm}
We kept the transmitter fixed and placed $9$ receivers along the trail at distances 
$50,53,56,\ldots,74$~meters respectively from the 
transmitter, and measured the mean received power (averaged over fading) at all receiving nodes. 
 We repeated this with 
varying the transmitter location $25$ times, thereby obtaining $25$ realizations of the network 
with $9$ links of length $50,53,56,\cdots,74$~meters in each realization (we chose the link lengths at least $50$~meters 
 because in reality 
the step size will be at least tens of meters). 
Under a given network realization, for 
the $i$-th link of length $r_i$~meters and shadowing realization $\nu_i$~dB, the mean received power 
 in dBm (averaged over fading; see Section~\ref{subsection:channel_model} for channel model):
\begin{equation}
\phi_{i}=\phi_{0}-10\eta\log\left(\frac{r_{i}}{r_{0}}\right)+\nu_{i}\:,1\leq i\leq 9\label{eqn:channel_equation_in_dBm}
\end{equation}
where $\phi_{0}$ is the mean received power (in dBm) at distance $r_0$.
%


\vspace{-0mm}
\subsubsection{Estimation of $\eta$, $\sigma$ and the shadowing decorrelation distance}
\label{subsubsec:estimating_eta_sigma}
\vspace{-1mm}
According to Gudmundson's model \cite{gudmundsonl-91correlation-model}, 
covariance between shadowing in two different links with one end fixed and the other ends on the same line at 
a distance $d$ from each other can be modeled by $ R_{X}(r_i,r_j)=\sigma^2 \exp(-d/D)$ where 
$\sigma$ denotes standard deviation (in dB) of shadowing random variables, and $D$ is a constant. 
Let $\mathbf{\theta}:=[\phi_0 \hspace{0.1in} \eta \hspace{0.1in} D\hspace{0.1in} \sigma^2]$. 
Define $ \nu_i^k $ to be the shadowing random variable for the link from the transmitter to node $i$ for the $k$-th realization 
of the network, where $1 \leq i \leq 9$ and $1\le k\le 25$. Assuming that  
$\mathbf{\nu^k}:=[ \nu_1^k\hspace{0.1in} \nu_2^k\ldots \nu_M^k]'$ is jointly Gaussian 
with covariance matrix denoted by $\mathbf{C(\theta)}$ (elements of this matrix are 
determined by Gudmundson's model), and $\mathbf{\nu^k}$ is i.i.d. 
across $k$, we calculate the maximum likelihood estimate $\hat{\theta}_{MLE}$: 
$\hat{D}_{MLE}=2.6$~meters, $\hat{\sigma}_{MLE}=7.7$~dB, $\hat{\eta}_{MLE}=4.7$. 
The correlation coefficient of shadowing between two links is less than $0.1$ beyond 
$2.3D$ distance, which implies that beyond $5.98$~meters the shadowing can be safely 
assumed to be independent. Hence, we need $\delta \geq 6$~m.

\vspace{2mm}
\subsubsection{Binary Hypothesis Based Approach to find the Shadowing Decorrelation Distance}
\label{subsubsec:hypothesis_test_D_normality}
\vspace{-2mm} The sample correlation coefficient $\hat{\rho}(r)$
between shadowing of all pairs of links whose transmitter is common
and the receivers are $r$ distance apart from each other is computed
as a function of $r$.  We want to decide whether the shadowing losses
over two links with a common receiver but whose transmitters are
separated by distance $r$ are correlated. Define the null Hypotheses
$H_{0}:\rho=0$ and the alternate Hypotheses $H_{1}:\rho\neq0$.  For a
target false alarm probability $\alpha=0.05$ (called the significance
level of the test), it turns out that we need $\hat{\rho}(r) \leq
0.34$, which requires $r \ge 3$~meters. {\em Hence, under the jointly
  lognormal shadowing assumption, shadowing is independent beyond
  $3$~meters.}


\vspace{-0mm}
\subsubsection{Testing Normality of Shadowing Random Variable via Non-Parametric Tests} 
\label{subsubsec:non_parametric_tests}
\vspace{-1mm} We picked $25$ links from $25$ independent network
realizations, and calculated their shadowing gains $\nu_i, 1 \leq i
\leq 25$ from (\ref{eqn:channel_equation_in_dBm}). Then we applied
Kolmogorov-Smirnov One Sample test (see
\cite{bickel-docksum01-statistics}): define the null hypothesis
$\mathcal{H}_{0}$ to be the event that the samples are coming from
$\mathcal{N}(0,\hat{\sigma}_{MLE}^2)$ distribution, and
$\mathcal{H}_{1}$ to be the event that they do not.  The test accepted
$H_0$ with level of significance  $0.05$. {\em Hence,
  lognormal shadowing is a good model in our setting.}

\vspace{-2mm}
\subsection{Number of packets to be transmitted for link evaluation}\label{subsection:coherence_time}
\vspace{-3mm}

In the experiments, in order to measure the outage probability of a
link, at a given transmit power a certain number of packets are sent
and their RSSI values recorded. To arrive at the required number of
packets we conducted the following experiment. Over several links in
the field, $5000$ packets were sent at intervals of $50$~ms, and their
RSSI values were recorded. We then characterise the coherence time of
the fading process by modeling it as a two state process.  We say that
the channel is in ``Bad'' state when the packet RSSI falls below the
mean RSSI (over packets) of the link by $20$~dB, otherwise the link is
in ``Good'' state. From the per-packet RSSI values in the $5000$
packet experiment, we observed that the mean number of packet
durations over which a channel remains in ``Good'' state is $56$,
i.e., $2.8$ seconds, and that the mean duration of the ``Bad'' state
is $100$~ms.  {\em Hence, we conclude that sending $2000$~packets
  ($100$~seconds duration, approximately $33$ Good-Bad cycles) is
  sufficient for the fading to be averaged out.}

  \begin{figure}[t!]
\begin{centering}
\begin{center}
\includegraphics[height=3cm, width=4.2cm]{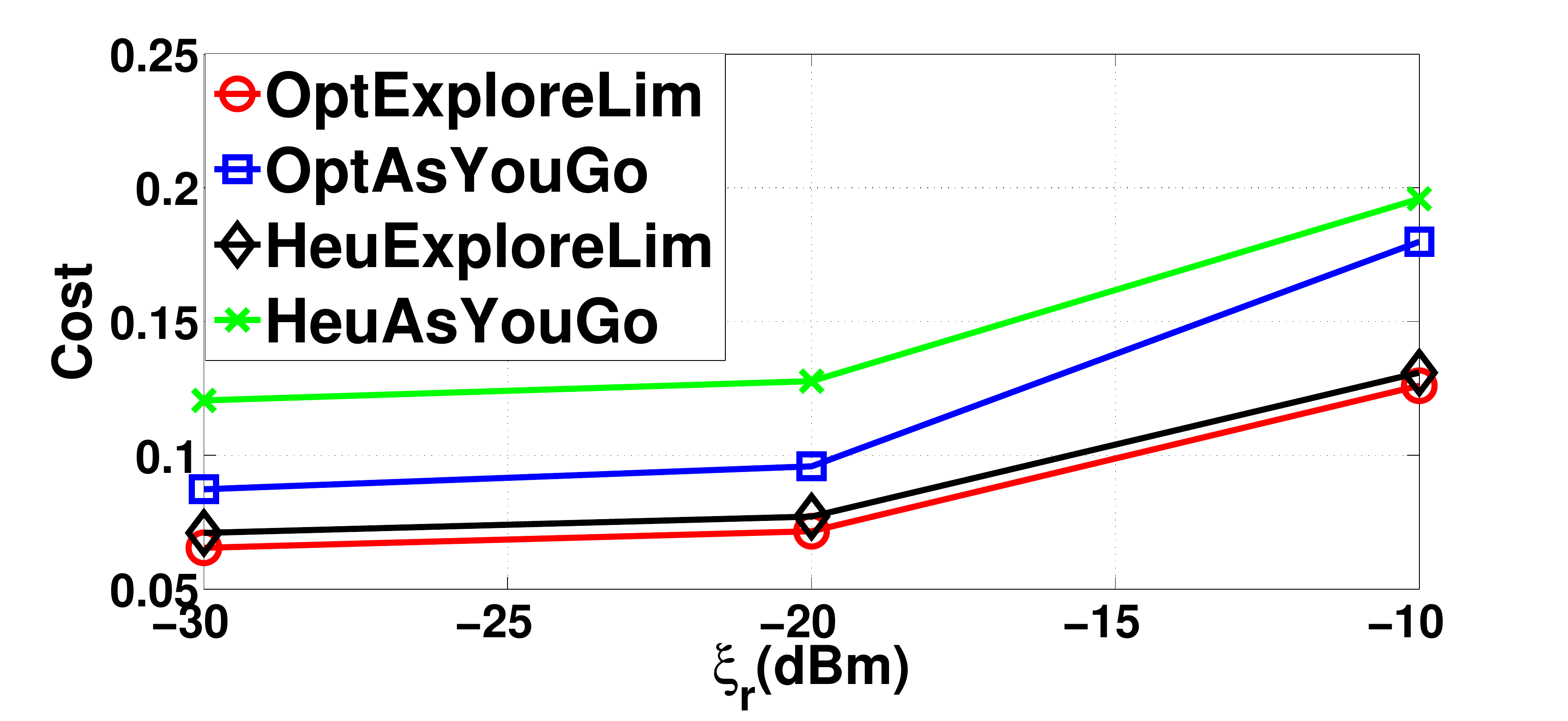}
\includegraphics[height=3cm, width=4.2cm]{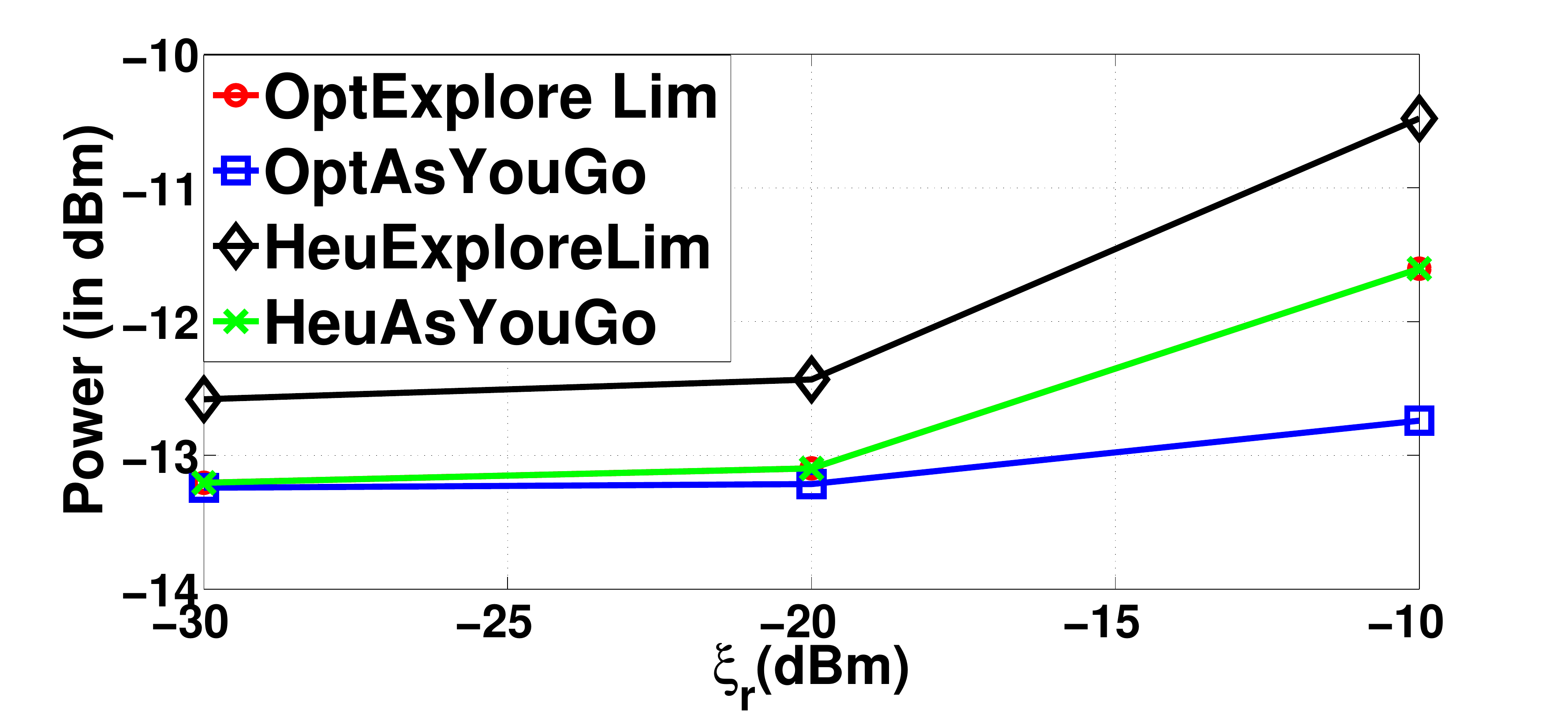}
\includegraphics[height=3cm, width=4.2cm]{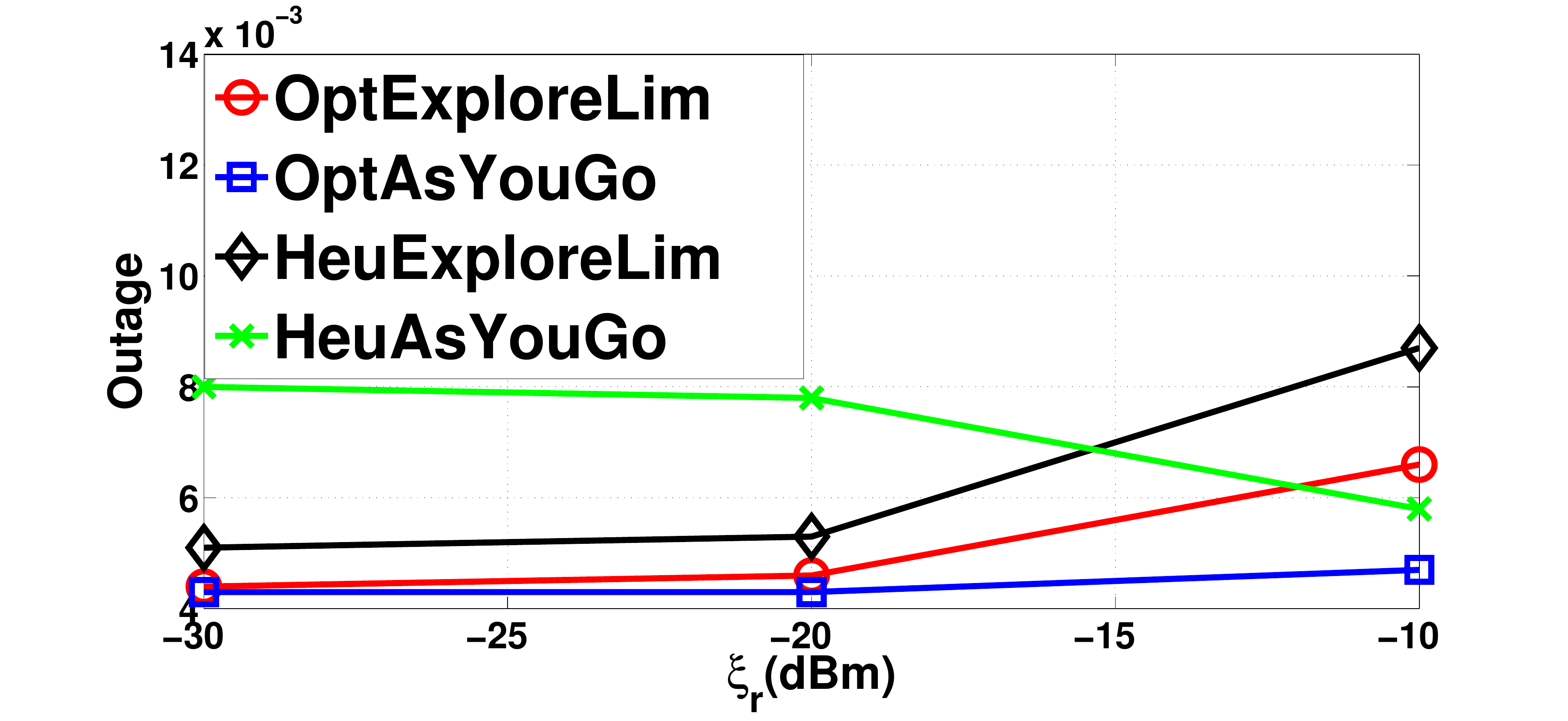}
\includegraphics[height=3cm, width=4.2cm]{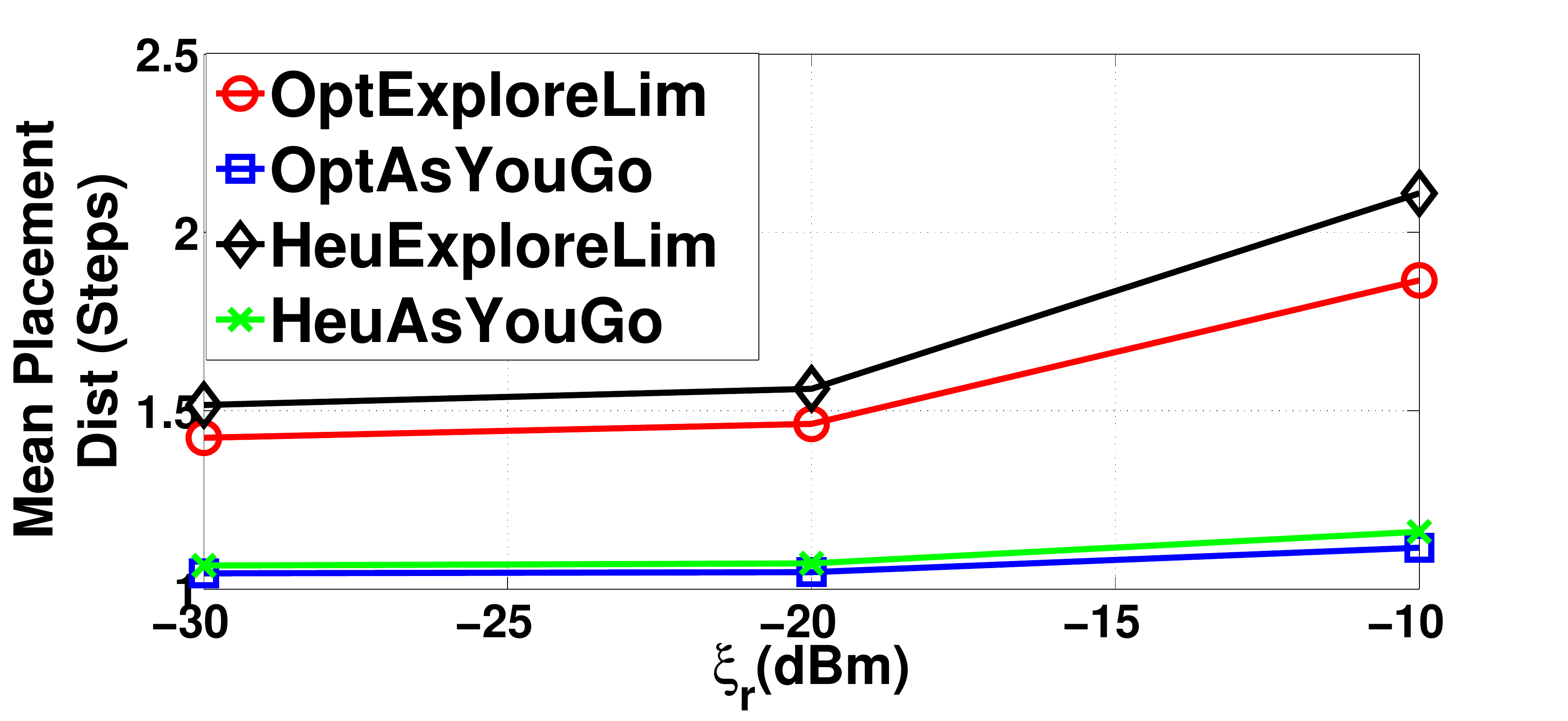}
\end{center}
\end{centering}
\vspace{-3mm}
\caption{Model-based results for $\xi_o=10$: mean cost per step, mean power per link, mean outage per link and 
mean placement distance (steps) vs. $\xi_r$ for the four algorithms: 
OptExploreLim, OptAsYouGo, HeuExploreLim, and HeuAsYouGo. 
The node power in the HeuAsYouGo algorithm was taken to be the same as the mean node power with the OptExploreLim algorithm. 
The unit of $\xi_r$ is actually mW, but here it is shown in dBm.}
\label{fig:cost-power-outage-distance-model-based_souryal_one_step_back}
\vspace{-5mm}
\end{figure}


\vspace{-3mm}
\section{Numerical and Experimental Results}\label{section:virtual_walking_experimental_results}
\vspace{-3mm}

In this section, we use TelosB motes with $9$~dBi antenna. 
The set of transmit powers $\mathcal{S}=\{-25,-15,-10,-5,0\}$~dBm, and outage is the
event $RSSI<-88$~dBm.
We used $\eta=4.7$, $\sigma=7.7$~dB (obtained from Section~\ref{subsection:shadowing_modeling}),  
and the step size $\delta=11$~m (since the shadowing decorrelation distance is $6$~m).

{\em Choice of $B$:} Define a link to be good if its outage probability
is less than $3\%$, and choose $B$ to be the largest integer
such that the probability of finding a good link of length $B \delta$
is more than $20\%$, under the highest transmit power. 
With $\eta=4.7$ and $\sigma=7.7$~dB, 
$B$ turned out to be $5$.

\vspace{-3mm}
\subsection{Observations from average cost per step estimates}\label{subsection:cost-comparison-algorithms}
\vspace{-3mm}
From $\eta$ and $\sigma$, we computed $\lambda^*$ and $c_{th}(r)$ 
(see Sections~\ref{subsection:opt-back-description} and \ref{subsection:opt-noback-description}) for various values of 
$\xi_o$ and $\xi_r$, 
and computed for each algorithm the mean cost per step, the mean outage probability per link, the mean length of a link 
and the mean power per link, assuming that the channel model is specified by 
the values of $\eta$ and $\sigma$ computed from the experiment. We call this approach the 
``model-based approach'' since we numerically compute the performance of the algorithms in an hypothetical 
homogeneous trail along which the propagation model is parameterized by the path 
loss exponent $\eta$ and the shadowing variance $\sigma$. {\em We keep the HeuAsYouGo  
transmit power and the per-link target outage equal to the mean power per link and mean outage per link of 
OptExploreLim.}  
We will only provide results for $\xi_o=10$; 
with this value the performance is satisfactory in terms of mean power per node, and the end-to-end outage 
for a network of length few hundreds of meters. 
The results are shown in 
Figure~\ref{fig:cost-power-outage-distance-model-based_souryal_one_step_back}.  
{\em Note that  performance of OptExploreLimLearning  is not shown in 
Figure~\ref{fig:cost-power-outage-distance-model-based_souryal_one_step_back}, since it  has the same asymptotic 
performance as OptExploreLim (since the policy in OptExploreLimLearning converges to the optimal policy with 
probability $1$).} 
\vspace{-1mm}
\subsubsection{Mean Placement Distance}
Pure as-you-go algorithms (OptAsYouGo, HeuAsYouGo of Figure~\ref{fig:cost-power-outage-distance-model-based_souryal_one_step_back}) 
place relays sooner than the algorithms that explore forward before placing a relay 
(OptExploreLim, HeuExploreLim). This is as expected, since they do not have the advantage of 
exploring over several locations and then picking the best. A pure as-you-go approach tends to be cautious, 
and therefore tries to avoid a high outage by placing relays frequently. 
As $\xi_r$ (cost of a relay) increases, relays will be placed less frequently.

\vspace{-1mm}
\subsubsection{Mean Power per Link}
Increasing $\xi_o$ (the cost per unit outage) will lower outage and 
      hence the transmit power increases. Increasing $\xi_r$ will place relays 
      less frequently, hence the transmit power increases. OptAsYouGo has smaller placement distance compared to 
      OptExploreLim and HeuExploreLim and hence it uses less power at each hop; we note, however, 
that OptAsYouGo places more relays, and, hence, could still end up using more total power.

      \vspace{-1mm}
\subsubsection{Mean Outage per Link}
As $\xi_o$, the penalty for outage,  increases, the mean outage per link decreases. 
As $\xi_r$ increases, the mean outage per link increases because 
we will place fewer relays with higher inter-relay distances. 
 HeuAsYouGo has outage 
probability comparable to other algorithms, but it pays in terms of number of relays since 
it places relays very frequently. 
We observe that the per-link outage decreases with $\xi_r$ for   HeuAsYouGo. As 
$\xi_r$ increases, the node power and the target outage (chosen from OptExploreLim) increases 
in such a way that the per-link outage decreases.  
\vspace{-1mm}
\subsubsection{Network Cost Per Step}
Cost increases with $\xi_r$ and $\xi_o$. 
 OptAsYouGo has a larger cost than OptExploreLim and HeuExploreLim, owing to shorter links. The cost of HeuAsYouGo 
      in Figure~\ref{fig:cost-power-outage-distance-model-based_souryal_one_step_back} is high due to smaller 
      placement distance. {\em Cost of HeuExploreLim  is very close to OptExploreLim. }

\vspace{-3mm}
\subsection{Deployment Experiments By ``Virtual'' Walking}\label{subsection:virtual_walking_result}

Now we report our results of carrying out an experimental evaluation of all the algorithms. 
Based on this evaluation, we select the best algorithm and suitable parameters in order to carry out an 
actual impromptu deployment, which we report in Section~\ref{sec:real_deployment}. 
Our experimental approach is the following: 
(i) We deploy $11$ TelosB motes, with $9$~dBi antennas, equally spaced by $11$~meters ($\delta=11$~m), 
lashed to trees along one edge of a $110$~meter trail, 
(ii) On command, one by one, each mote broadcasts $2000$~packets at each power level from the set   
$\mathcal{S}=\{-25,-15,-10,-5,0\}$~dBm, while the rest remain in the receive mode. 
For each transmit power level of a node, the outage probability at each other receiving node is
recorded.

Next, we applied the policies to the data.  Since we have gathered the outage
probabilities of every possible link for all power levels, we have all
possible measurements that can be possibly made during an actual
deployment.  Thus, we can use the measurements to determine the actual
network that will be deployed if an agent was to walk along the trail
starting from sink at location 1, with the source being at location 11 (the
distance between the sensor and the base station is $110$~meters).  We
choose $\xi_o=10$ and $\xi_r=0.01$ for the virtual deployment.  For the HeuAsYouGo algorithm, we randomized
between two power levels from $\mathcal{S}$ so that the mean transmit power
per node in the data-based HeuAsYouGo remains equal to that of
model-based OptExploreLim. We have also calculated the optimal
end-to-end cost of the network graph; we calculated the shortest path
from node $11$ to node $1$ over the weighted network graph where the
weight of any potential link consists of the transmit power, outage
cost and relay cost, and the weights are available from the field
measurements. The cost of the sensor node is not taken into account.
Obviously, this approach (OptExploreAll) gives the
smallest end-to-end cost 
 (see Table~\ref{table:virtual_deployment_comparison1} and the abbreviations in
its caption).  If we place a relay at location $i$, ($i
\geq 7$), the remaining number of steps from there to the sensor
location is less than $B=5$. In that case, we place one more relay
between locations $i$ and $11$ such that the cost between $11$ and $i$
is minimized. {\em For OptExploreLimLearning, we
  chose $\lambda_0=0.0321$ (optimal cost per step
  when $\eta=4$, $\sigma=7$~dB).}

\begin{table}[t!]
\footnotesize
\centering
\begin{tabular}{|c |c |c |c |c |c |}
\hline
Algo- & Relay & No. of mea- & Total Po-
 & Sum  & Total  \\ 

 rithm   &  location&  surements & wer (mW) & Outage & Cost \\
\hline
OEL & 5,7,9 & 17 & 0.3542 & 0.004 & 0.424\\
\hline
HEL & 5,7,9 & 17 & 0.3542 & 0.004 & 0.424 \\
\hline
OELL & 4,8,10 & 15 & 0.3826 & 0.006 & 0.472 \\
\hline
OAYG & 2,3,4,5,6,8,9 & 10 & 0.747 & 0.018 & 0.997 \\
\hline
HAYG & 2,3,5,6,7, & 10 & 0.451 & 0.586 & 6.391 \\
     &     8,9,10             &     &      &        &      \\
\hline
OEA & 2,6,9 & 40 & 0.0704 & 0.002 & 0.1204 \\
\hline
\end{tabular}
\vspace{-0mm}
\caption{Results of virtual walking deployment for one side of a trail, for $\xi_r=0.01$ and $\xi_o=10$. 
Abbreviations: OEL-OptExploreLim, 
HEL-HeuExploreLim, OELL-OptExploreLimLearning, OAYG-OptAsYouGo, HAYG-HeuAsYouGo, OEA-OptExploreAll.}
\vspace{-0mm}
\label{table:virtual_deployment_comparison1}
\vspace{-0mm}
\end{table}
\normalsize
\vspace{-0mm}

The OptExploreLim algorithm places relays at locations 5,7,9.  It
requires $B=5$ measurements for placing at each of $5$ and $7$, see
Table~\ref{table:virtual_deployment_comparison1}. Then it has to place
one more relay, for that it has to measure the cost of the following
paths: $\{(11,10),(10,7)\}$, $\{(11,9),(9,7)\}$, $\{(11,8),(8,7)\}$
and $(7,7)$. Hence, OptExploreLim requires $17$ measurements to place
3 relays. Similar logic follows for HeuExploreLim and OptExploreLimLearning.  The as-you-go
algorithms require $10$ measurements for $10$~steps, and place 7-8
relays.  Exploration algorithms place a smaller number of relays and
yet produce much better performance compared to pure
as-you-go. OptExploreAll (the best algorithm) significantly improves the performance
compared to exploration algorithms, but makes $40$ measurements (between all
potential location pairs whose distance is less than or equal to $5$~steps).  
As-you-go algorithms work with measurements acquired
as the agent walks and, hence, are suboptimal. {\em Hence, the algorithms that employ 
partial exploration are the ones that  require reasonable number
of measurements while giving satisfactory performance.}

\begin{figure}[!t]
\centering
\begin{center}
\includegraphics[height=2.8cm, width=4.9cm]{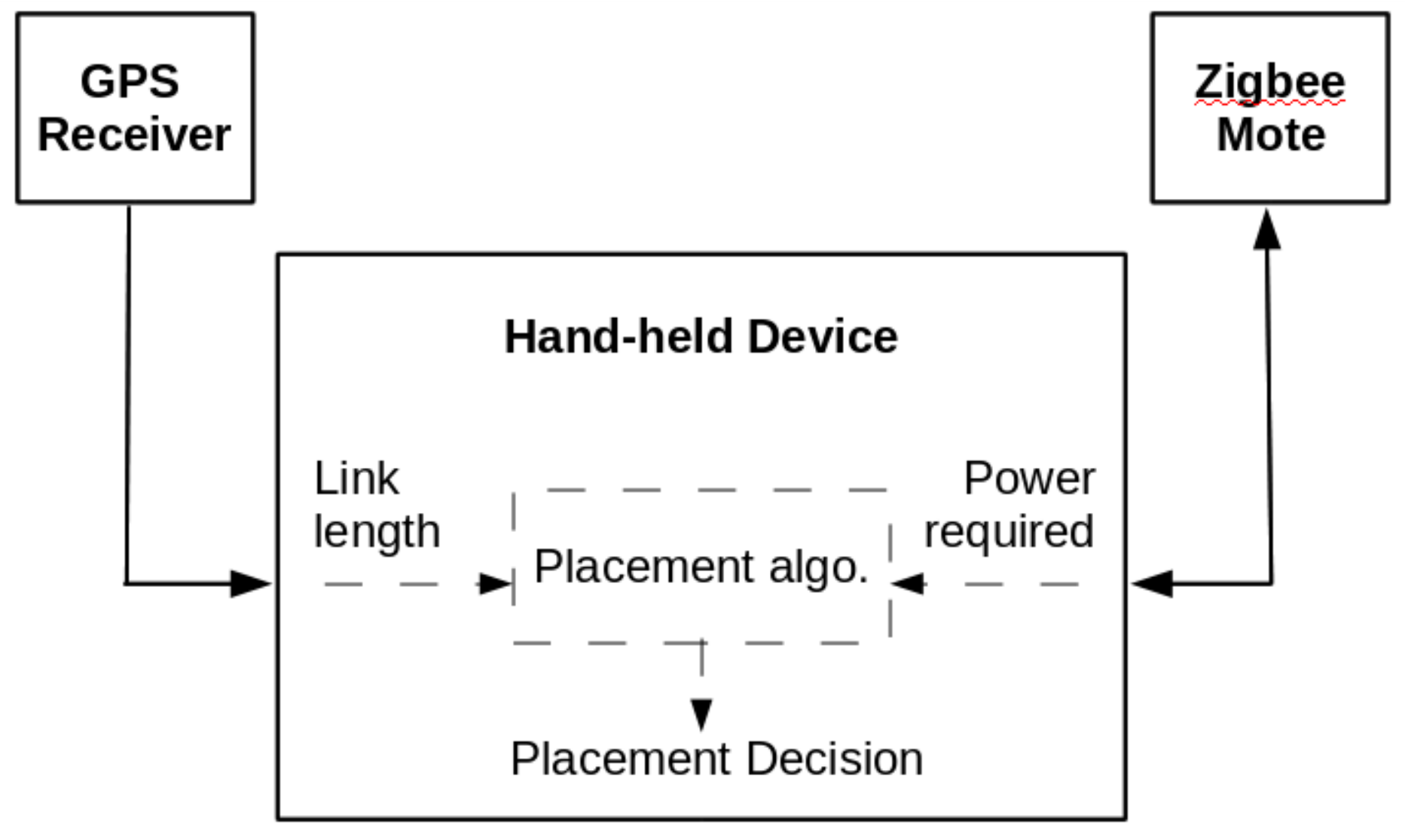}
\includegraphics[height=2.8cm, width=3.5cm]{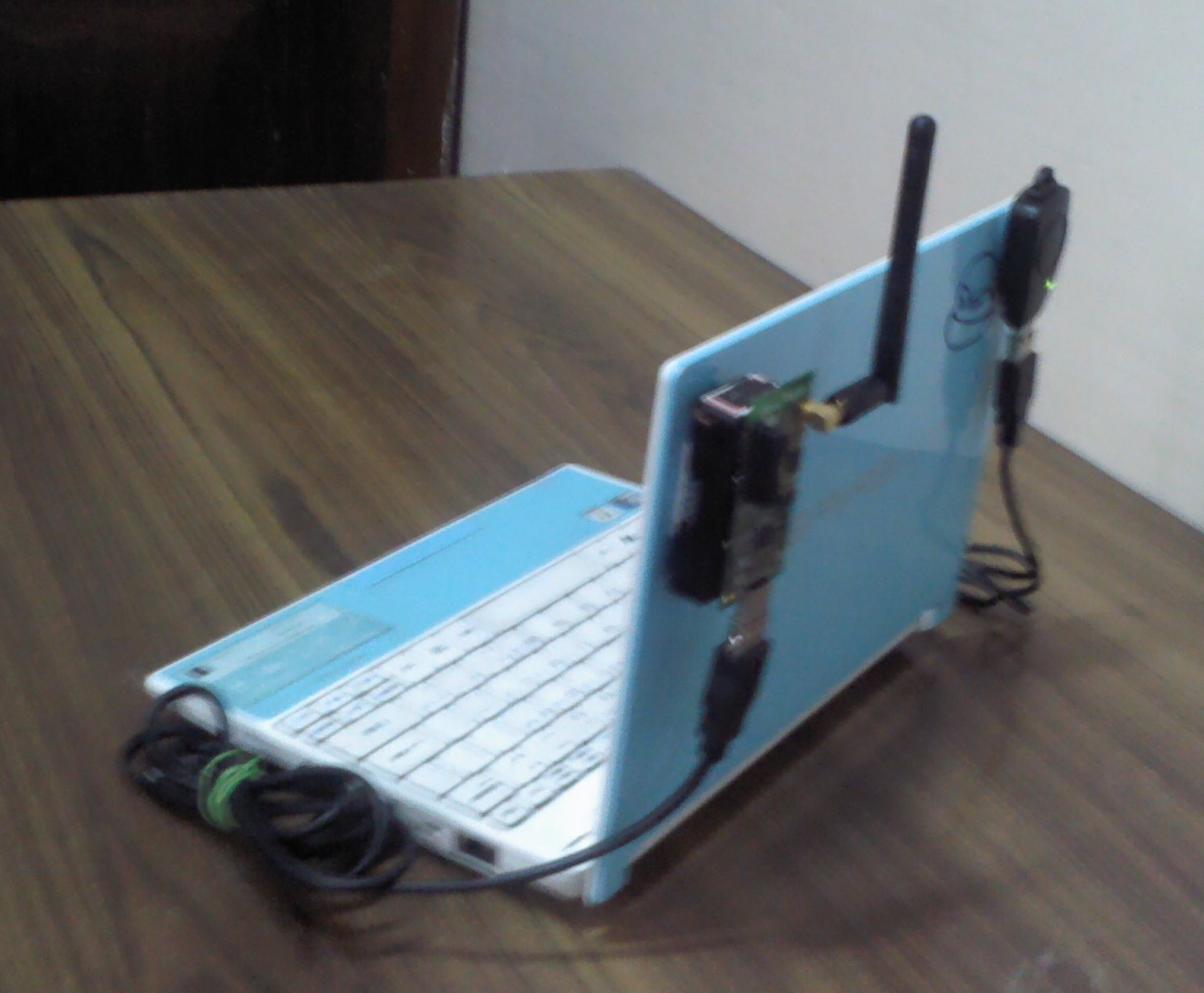}
\end{center}
\vspace{-4mm}
\caption{Architecture of the deployment tool (basically a notebook computer with a 
GPS dongle and a Zigbee mote (the commanding node (CN); see text) attached via a USB port).}
\label{fig_3:Architecture_of_deployment_tool}
\vspace{-0mm}
\end{figure}

\vspace{-2mm}
\section{Implementation for Actual Deployment}\label{section:implementation}
\vspace{-2mm} 

In Section~\ref{subsection:virtual_walking_result} we
provided results from "virtual walking" deployments which are
basically off-line computations that utilised detailed field
measurements of all possible potential links in the field. We next 
turn to our experiments with "actual walking" deployments.  The
deployment agent carrying a deployment tool (see
Figure~\ref{fig_3:Architecture_of_deployment_tool}) goes about
executing the process described here starting from the base station
and proceeds all the way to the location where the sensor has to be
placed.  On the way, the relay nodes are placed as guided by the
placement algoithm running on the deployment tool. The previously
Placed Node (PPN) is the most recently placed node in the
deployment. The evaluation node (EN) is the node that is evaluated by
the placement algorithm; it is lashed to a tree at (or just near) a
potential placement point (i.e., a multiple of $\delta$ steps from the
PPN).  The commanding node (CN) is the command relay node which is
attached to the deployment tool from which the deployment agent will
issue commands for evaluation. At a potential location, the agent
issues a command to the EN via CN to evaluate the link between EN and
PPN.  After performing link quality measurements such as link outage,
at different transmit power levels, the EN reports the results to the
CN. After evaluating the requisite number of EN locations (one for
pure as-you-go deployment, and $B$ for exploration based algorithms),
the placement algorithm decides where the EN should be placed.  The
procedure is repeated until the source location is
reached. 

The mote side of the deployment code is written in TinyOS and the node
placement code uses C and Matlab. In hardware, the deployment tool
consists of a USB GPS receiver, a mote, and a handheld device (netbook
running Fedora 12 distribution) that can interface with the GPS and
the mote (Figure \ref{fig_3:Architecture_of_deployment_tool}).  We
have used a low cost off-the-shelf GPS USB receiver dongle for getting
the location information. This receiver is based on the SiRF STAR III
chipset. In our GPS experiments, we got an error of roughly 5 meters
for less than 50 meters distance in the best case where there is a
clear sky and good satellite visibility. Considering that practical
deployment distances would be much higher than 50 meters (particularly
with high power motes), we found that the GPS error is within the
acceptable limits.

\begin{figure}[t!]
\begin{centering}
\includegraphics[height=3.3cm, width=9cm]{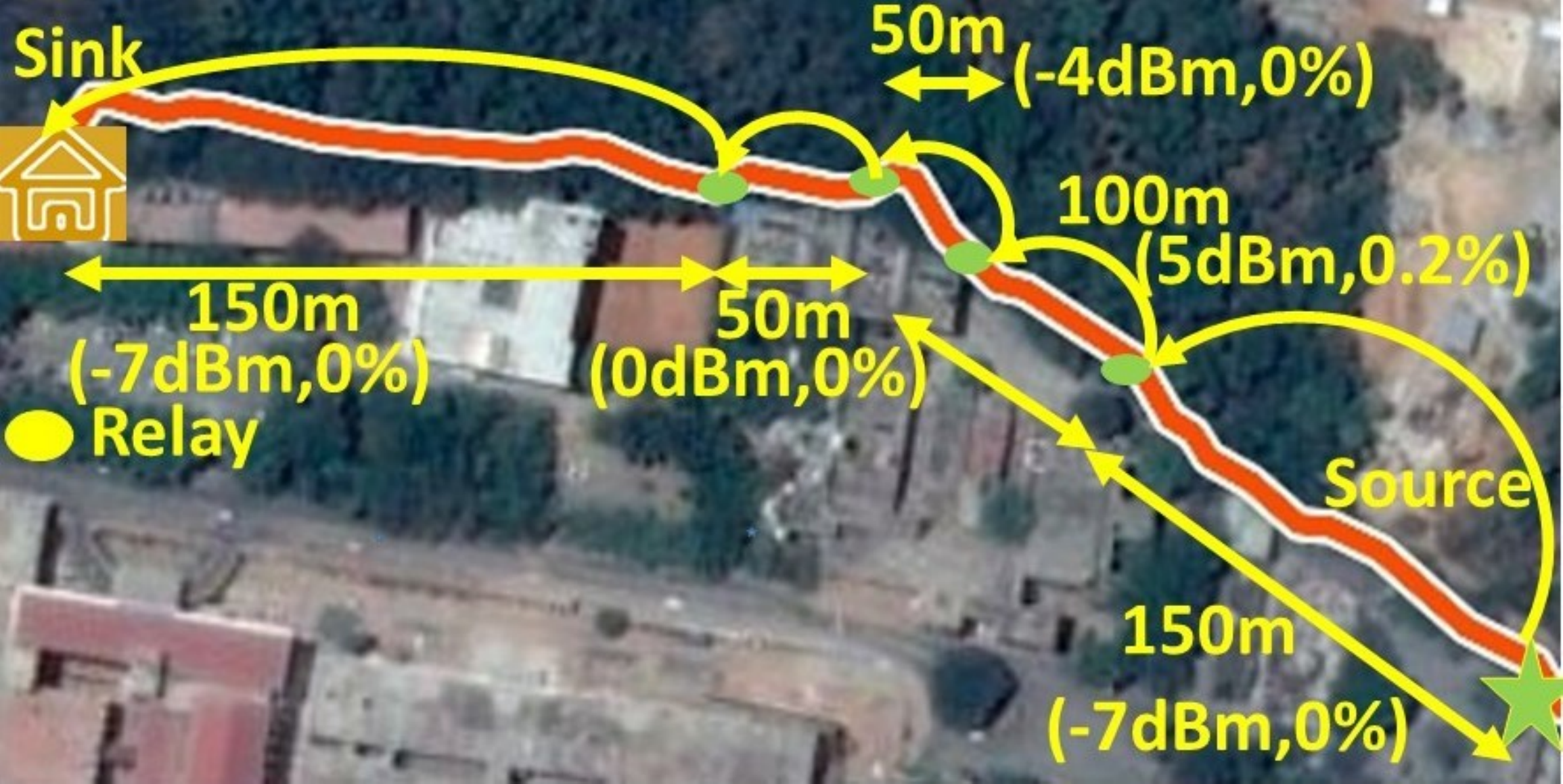}
\includegraphics[height=2.5cm, width=9cm]{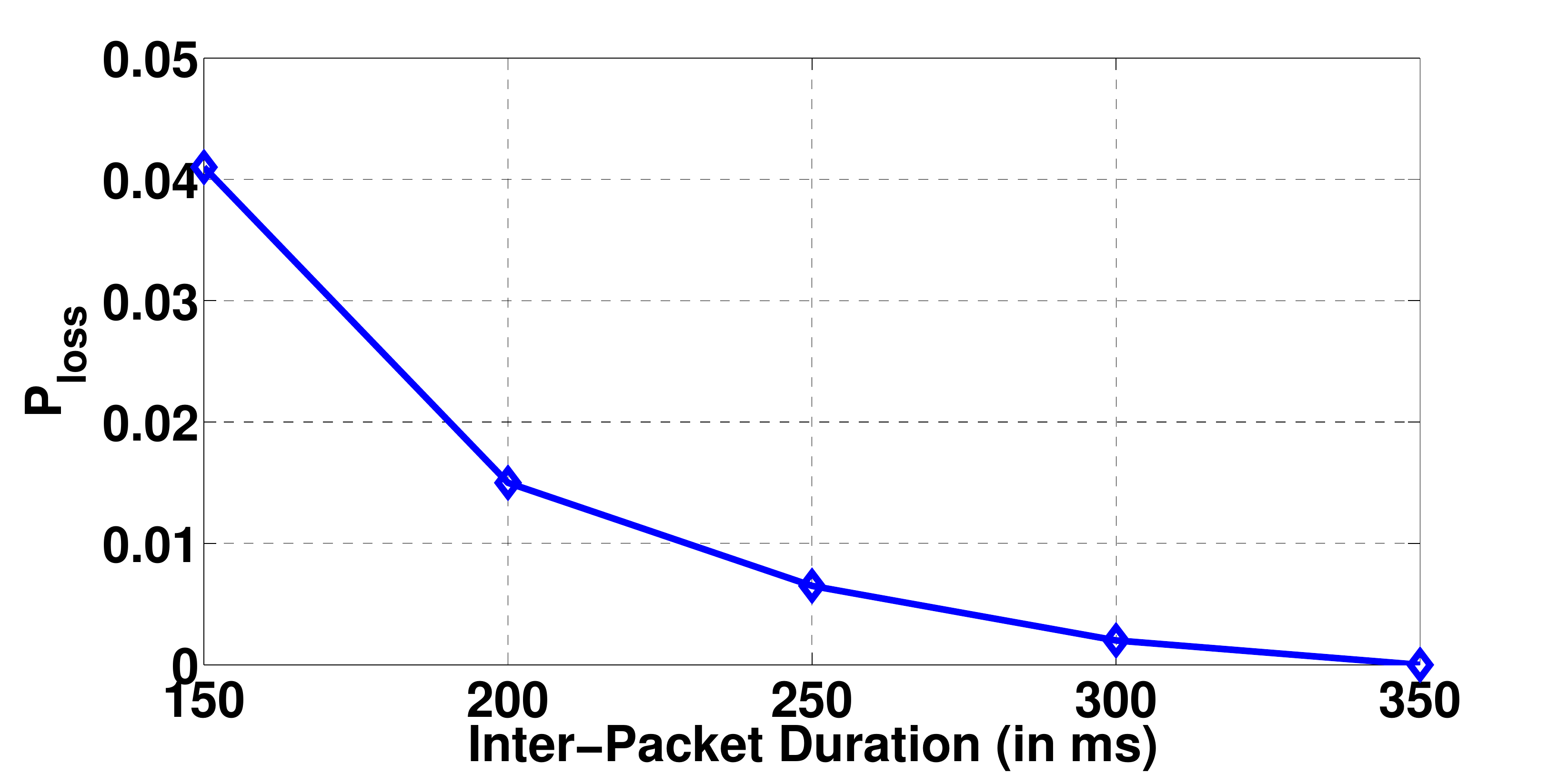}
\end{centering}
\vspace{-0mm}
\caption{Real deployment along a long trail using OptExploreLimLearning  with iWiSe motes, $\xi_o=100$, $\xi_r=1$: 
five nodes (including the source) are placed; link lengths, transmit powers, and $\%$ outage 
probabilities are shown; the plot shows variation of end-to-end loss probability with 
inter-packet duration for periodically generated packets from the source}
\label{fig:real_deployment_OptExploreLimLearning}
\vspace{-0mm}
\end{figure}

\vspace{-5mm}
\section{Physical Deployment Experiments}\label{sec:real_deployment}
\vspace{-4mm}  
With the experience (on the choice of deployment algorithm) obtained from the virtual 
deployment experiment discussed in Section~\ref{section:virtual_walking_experimental_results}, 
we performed some real deployment experiments along
a long trail in our campus (not exactly linear in shape, which is the reality in a practical forest environment) with 
the powerful iWiSe motes (see \cite{iwise})  
equipped with $9$~dBi antennas\footnote{``Actual walking" deployment 
was done using the deployment tool.}. 
We chose $\xi_o=100$, $\xi_r=1$, $B=5$~steps, $\delta=50$~meters, and  $\mathcal{S}=\{-7,-4,0,5\}$~dBm.  
$P_{rcv-min}=-97$~dBm; the PER at this RSSI becomes $2\%$ for iWiSe motes
(obtained experimentally). In this deployment experiment we used the 
PER of a link as a substitute for outage probability; this does not violate the 
basic assumptions of our formulation, and the algorithms remain the same. For  $\eta=4$, 
$\sigma=7$~dB, the optimal average cost per step is $1.0924$
(computed numerically using policy iteration). 
Taking $\lambda_0=1.0924$ (the initial {\em guess}), we performed real deployment experiments  with
OptExploreLimLearning.  The
deployed network is shown in Figure~\ref{fig:real_deployment_OptExploreLimLearning}. 
The  sink is denoted by the ``house'' symbol. 
{\em The two short ($50$~m long) links account for significant path-loss due to the turn in the trail.} 
After completion of the deployment,
we used the last placed node as the source and sent periodic traffic from the source to the sink node at 
various rates. As the arrival rate
increases, the loss probability increases (see Figure~\ref{fig:real_deployment_OptExploreLimLearning}) due to carrier sense
failures and collisions because of simultaneous transmissions
from different nodes.  For very low arrival rate, the loss
probability becomes $0$ even though the sum PER under the lone
packet model is not $0$. This happens because of
link level retransmissions and the relatively short outage
durations; even if a packet encounters an outage on a link
along the path, retransmission attempts succeed with high probability. 
We see that, even though the design was for the lone packet model, the network can carry 
$4$~packets/second with $P_{loss} \leq 1\%$.

\vspace{-4mm}
\section{Conclusion and Ongoing Work}\label{sec:conclusion}
\vspace{-3mm} In this paper, we have compared the on-field performance
of several as-you-go deployment algorithms. Pure as-you-go networks
need to be overly cautious, and, hence, deploy far too many
relays. Our results suggest that limited exploration based on-line
algorithms (such as HeuExploreLim, and OptExploreLimLearning) provide
satisfactory performance, at the cost of some additional measurements.
In a large forest, we can deploy
using OptExploreLimLearning in one trail and use the updated policy in
another trail so that the per-step cost of the entire network is optimal. There are several issues left for
future study: (i) robust deployment against seasonal variation of
propagation,  and (ii)  deployment in 2D and 3D regions.

\vspace{-5mm}

\renewcommand{\thesubsection}{\Alph{subsection}}

\appendices

\bibliographystyle{unsrt}
\bibliography{chattopadhyay-etal14deployment-forest-trail}

\begin{thebibliography}{10}

\bibitem{chattopadhyay-etal13backtracking-as-you-go_arxiv}
A.~Chattopadhyay, M.~Coupechoux, and A.~Kumar.
\newblock As-you-go deployment of a wireless network with on-line measurements
  and backtracking.
\newblock {\em http://arxiv.org/abs/1308.0686}.

\bibitem{chattopadhyay-etal13measurement-based-impromptu-placement_wiopt}
A.~Chattopadhyay, M.~Coupechoux, and A.~Kumar.
\newblock Measurement based impromptu deployment of a multi-hop wireless relay
  network.
\newblock In {\em Proc. of the 11th Intl. Symposium on Modeling and
  Optimization in Mobile, Ad Hoc, and Wireless Networks (WiOpt)}. IEEE, 2013.

\bibitem{souryal-etal07real-time-deployment-range-extension}
M.R. Souryal, J.~Geissbuehler, L.E. Miller, and N.~Moayeri.
\newblock Real-time deployment of multihop relays for range extension.
\newblock In {\em Proc. of the ACM International Conference on Mobile Systems,
  Applications and Services (MobiSys), San Juan, Puerto Rico, June 2007}, pages
  85--98. ACM, 2007.

\bibitem{liu-etal10breadcrumb}
H.~Liu, J.~Li, Z.~Xie, S.~Lin, K.~Whitehouse, J.A. Stankovic, and D.~Siu.
\newblock Automatic and robust breadcrumb system deployment for indoor
  firefighter applications.
\newblock In {\em Proc. of the ACM International Conference on Mobile Systems,
  Applications and Services (MobiSys)}, 2010.

\bibitem{aurisch-tlle09relay-placement-emergency-response}
T.~Aurisch and J.~T\"{o}lle.
\newblock Relay {P}lacement for {A}d-hoc {N}etworks in {C}risis and {E}mergency
  {S}cenarios.
\newblock In {\em Proc. of the Information Systems and Technology Panel (IST)
  Symposium}. NATO Science and Technology Organization, 2009.

\bibitem{howard-etal02incremental-self-deployment-algorithm-mobile-sensor-netw%
ork}
M.~Howard, M.J. Matari\'{c}, and S.~Sukhat~Gaurav.
\newblock An incremental self-deployment algorithm for mobile sensor networks.
\newblock {\em Kluwer Autonomous Robots}, 13(2):113--126, 2002.

\bibitem{sinha-etal12optimal-sequential-relay-placement-random-lattice-path}
A.~Sinha, A.~Chattopadhyay, K.P. Naveen, M.~Coupechoux, and A.~Kumar.
\newblock Optimal sequential wireless relay placement on a random lattice path.
\newblock {\em Ad Hoc Networks Journal (Elsevier).}, 21:1--17, 2014.

\bibitem{bhattacharya-etal13smartconnect-comsnets}
A.~Bhattacharya, A.~Rao, D.~G.~Rao Sahib, A.~Mallya, S.M. Ladwa, R.~Srivastava,
  S.V.R. Anand, and A.~Kumar.
\newblock Smartconnect: A system for the design and deployment of wireless
  sensor networks.
\newblock In {\em Proc. of the 5th International Conference on Communication
  Systems and Networks (COMSNETS)}. IEEE, 2013.

\bibitem{bhattacharya-kumar12qos-aware-survivable-network-design}
A.~Bhattacharya and A.~Kumar.
\newblock Qo{S} aware and survivable network design for planned wireless sensor
  networks.
\newblock {\em http://arxiv.org/abs/1110.4746}.

\bibitem{gudmundsonl-91correlation-model}
M.~Gudmundson.
\newblock Correlation model for shadow fading in mobile radio systems.
\newblock In {\em Electronics letters}, volume~27, pages 2145--2146. IET, 1991.

\bibitem{bickel-docksum01-statistics}
P.J. Bickel and K.A. Doksum.
\newblock {\em Mathematical Statistics, volume I.}
\newblock Prentice Hall Englewood Cliffs, NJ, 2001.

\bibitem{iwise}
{\em \url{http://www.astec.org.in/astec/content/wireless-sensor-network},}.

\end{thebibliography}

\end{document}